# STATUS OF NB₃SN ACCELERATOR MAGNET R&D AT FERMILAB*

A.V. Zlobin†, FERMILAB, Batavia, IL 60510, U.S.A.

*Abstract*

New accelerator magnet technology based on Nb$_3$Sn superconductor is being developed at Fermilab since late 90's. Six short dipole models, seven short quadrupole models and numerous individual dipole and quadrupole coils have been built and tested, demonstrating magnet performance parameters and their reproducibility. The technology scale up program has built and tested several dipole and quadrupole coils up to 4-m long. The results of this work are summarized in the paper.

## INTRODUCTION

Dipole magnets for the LHC energy upgrade scenario with operating field of ~20 T would require using high-field high-temperature superconductors such as BSCCO or YBCO, which have highest upper critical magnetic field $B_{c2}$. However, due to the substantially higher cost and lower critical current density in magnetic fields below 15 T, a hybrid approach with Nb$_3$Sn superconductor in fields below 15 T is a quite attractive option even though the Nb$_3$Sn and HTS materials require different coil fabrication techniques.

During the past decade, Fermilab has been developing new Nb$_3$Sn accelerator magnet technologies in the framework of the High Field Magnet (HFM) program. Nb$_3$Sn accelerator magnets can provide operating fields up to 15 T and significantly increase the coil temperature margin. Such magnets are being developed for the LHC IR upgrade, Muon Collider Storage Ring, and present and future high-energy hadron colliders. The program began in 1998 with the development of the small-aperture arc dipoles for the Very Large Hadron Collider (VLHC) [1]. Since 2003, the emphasis of the program was shifted toward large-aperture Nb$_3$Sn quadrupoles for an LHC IR upgrade [2].

The High Field Magnet R&D program started with the development of basic technologies and studies of main magnet parameters (maximum field, quench performance, field quality) and their reproducibility using a series of short models, and then proceeded with the demonstration of technology scale up using relatively long coils. Along the way, the HFM program has made several breakthroughs in Nb$_3$Sn accelerator magnet technologies. The most important of them include the development and demonstration of high-performance Nb$_3$Sn strands and cables, reliable and reproducible coil fabrication technology, and a variety of accelerator quality mechanical structures and coil pre-load techniques. The status and the main results of the Nb$_3$Sn accelerator magnet R&D at Fermilab are summarized in this paper.

___________________________________________
* Work supported by Fermi Research Alliance, LLC, under contract No. DE-AC02-07CH11359 with the U.S. Department of Energy.
† zlobin@fnal.gov

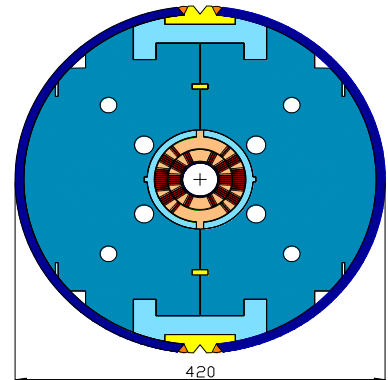

Figure 1: HFDA dipole cross-section.

## MAGNET DESIGNS AND PARAMETERS

### Dipole and quadrupole models

The design and main parameters of Fermilab's dipole models of the HFDA series are described in [3]. These magnets have been developed as baseline dipoles for the VLHC which was extensively studied in the U.S. around 2000 [4]. The cross-section of the dipole cold mass is shown in Fig. 1. This magnet was designed to provide a nominal field of 10-11 T ($B_{max}$~12 T) in a 43.5 mm aperture at an operating temperature of 4.5 K. The main R&D goal of this model magnet series was to develop robust Nb$_3$Sn coil technology and an inexpensive mechanical structure suitable for industrialization. This goal dictated the philosophy of magnet design and technology. The magnet design is based on a two-layer shell-type coil and a cold iron yoke. To reduce the magnet cost, a compact collarless mechanical structure with Al clamps, a 400 mm iron yoke and a 10 mm stainless steel skin was used.

The design and parameters of Fermilab's quadrupole models of TQC series are described in [5]. These magnets were proposed and used as a technological model of a new generation of large-aperture IR quadrupoles being developed by the US-LARP collaboration [6] for the planned LHC luminosity upgrade. The TQC cross-section is shown in Fig. 2.

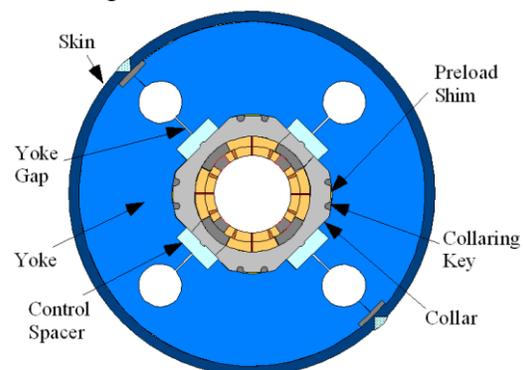

Figure 2: TQC quadrupole cross-section.

This model magnet series was designed to provide the same nominal field gradient of 200 T/m ($G_{max}$~250 T/m) in 90-mm aperture at the same operating temperature of 1.9 K as the present 70-mm NbTi IR quadrupoles (MQXB). The quadrupole design consists of a two-layer shell-type coil and a cold iron yoke. The design and technology of quadrupole coils used in TQC models largely rested on the results of the dipole program described above. The TQC quadrupole mechanical structure is based on the slightly modified mechanical structure of the present LHC IR quadrupoles (MQXB). It includes a 25-mm-thick round stainless-steel collar, a 400 mm iron yoke and a 12 mm thick stainless steel skin.

The dipole and quadrupole coils were wound using keystoned $Nb_3Sn$ Rutherford cables with 27 (28 in first dipole models) strands 0.7 mm (TQC) and 1.0 mm (HFDA) in diameter. The cable used in the first two dipole models HFDA02-03 had 0.025 mm thick stainless steel core to control the strand crossover resistance while the cables used in HFDA04-07 dipoles and in all the TQC quadrupole models were without a core. The dipole and quadrupole cable parameters are summarized in Table 1.

Table 1: Cable parameters.

| Parameter | Unit | HFDA | TQC |
|---|---|---|---|
| Number of strands |  | 27(28) | 27 |
| Strand diameter | mm | 1.00 | 0.70 |
| Cable mean thickness | mm | 1.80 | 1.26 |
| Cable width | mm | 14.24 | 10.05 |
| Cable keystone angle | deg | 0.9 | 1.0 |
| Cable insulation thickness | mm | 0.25 | 0.125 |

The design quench parameters at the corresponding nominal operating temperatures for the dipoles and quadrupoles, calculated for the strand critical current density $J_c$(12 T, 4.2 K)=2 kA/mm$^2$, are summarized in Table 2. Both magnets are designed for practically the same level of maximum field in the coil $B_{max}$~12 T at the corresponding nominal operating temperatures.

Table 2: Magnet quench parameters at 4.5 and 1.9 K.

| Parameter | Operating temperature | HFDA | TQC |
|---|---|---|---|
| $B_{max}$, T | 4.5 K | 12.05 |  |
| Quench current, kA |  | 21.66 |  |
| Coil peak field, T |  | 12.6 |  |
| $G_{max}$, T/m | 1.9 K |  | 233 |
| Quench current, kA |  |  | 14.07 |
| Coil peak field, T |  |  | 12.1 |

### $Nb_3Sn$ strands

Three types of $Nb_3Sn$ strand were used in the dipole and quadrupole model magnets.

The strand for the first three dipole models, HFDA02-04, was produced using the Modified Jelly Roll (MJR) process and had 54 $Nb_3Sn$ sub-elements in cross-section. The MJR strand had a critical current density $J_c$(12 T, 4.2 K)~2.0-2.2 kA/mm$^2$ and a quite large filament size $d_{eff}$~100 μm in 1-mm strand [7].

The strand for the last three dipole models, HFDA05-07, was made using the Powder-in-Tube (PIT) process and had 192 $Nb_3Sn$ filaments. The PIT strand had lower $J_c$(12 T, 4.2 K)~1.6-1.8 kA/mm$^2$ and smaller $d_{eff}$~50 μm at 1-mm strand diameter [7].

Then a new improved strand based on the Restack Rod Process (RRP) was developed [8]. This strand was initially produced with a 54/61 cross-section design and a high $J_c$(12 T, 4.2 K) up to 3 kA/mm$^2$. However, the quadrupole models TQC01 (a and b) and TQC02b were made using the MJR strand with lower $J_c$(12 T, 4.2 K)~2 kA/mm$^2$ and 54 sub-elements ($d_{eff}$~70 μm in 0.7-mm strand). The second generation of quadrupole models TQC02a, TQC02E (a and b) used the RRP strand with $J_c$(12 T, 4.2 K)~2.8 kA/mm$^2$ and 54 sub-elements.

Taking into account the importance of the strand and cable designs and parameters for accelerator magnet performance, an extensive $Nb_3Sn$ strand and cable R&D study was conducted by Fermilab in parallel with the model magnet R&D program focusing on the improvement of strand stability, reduction of strand magnetization, minimization of strand degradation during cabling, etc. RRP strands with various cross-section designs were produced and studied in collaboration with OST [9]. Based on the results of these studies, the RRP-108/127 strand with increased sub-element spacing and reduced sub-element size was developed as a baseline conductor for the $Nb_3Sn$ accelerator magnet R&D. This strand was used in several dipole and quadrupole coils.

The cross-sections of some $Nb_3Sn$ strands used in the dipole and quadrupole models are shown in Fig. 3.

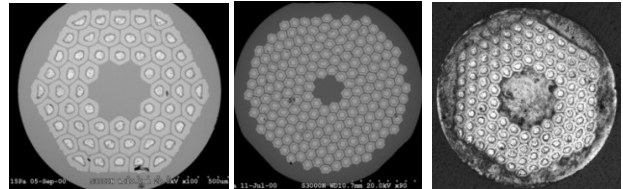

Figure 3: $Nb_3Sn$ strand cross-sections: a) MJR-54/61, b) PIT-192 and c) RRP-108/127.

### Cable insulation

Several types of cable insulation based on ceramic, S2-glass and E-glass fiber were studied [10] and used in the $Nb_3Sn$ dipoles and quadrupoles. The insulation types, dimensions and their costs are shown in Fig. 4.

The most important differences between these materials include mechanical and electrical strength after reaction, thicknesses, and cost. Ceramic insulation has demonstrated the best electrical strength and mechanical properties during coil processing. However, its thickness is relatively large and it is much more expensive than either the S2-glass or E-glass systems. The E-glass tape is the least expensive and most readily available in a variety of thicknesses, and based on tests is acceptable for use in $Nb_3Sn$ magnets (at least during an R&D phase).

All the dipole models were made using cables insulated with two-layers of the ceramic tape. The quadrupole models were made using cables insulated with the S2-glass sleeve. Some dipole and quadrupole coils were made using S2- or E-glass tapes or their combinations.

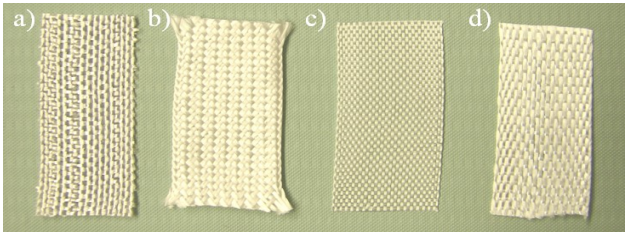

Figure 4: Insulation types and cost: a) 0.125 × 13 mm$^2$ ceramic tape (~20 \$/m), b) 0.125 mm S2-glass sleeve (~10 \$/m), c) 0.075 × 13 mm$^2$ E-glass tape (~0.2 \$/m), and d) 0.125 × 13 mm$^2$ S2/E-glass combination tape (~6 \$/m).

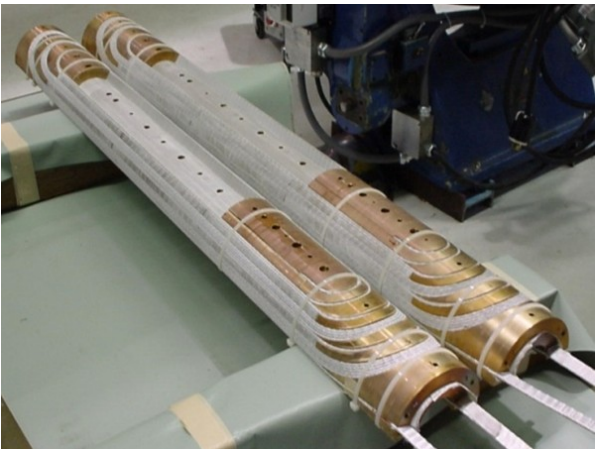

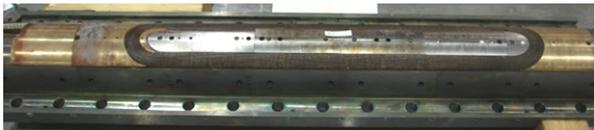

Figure 5: Nb$_3$Sn dipole coils impregnated with ceramic binder after winding and curing (top) and a quadrupole coil cured with ceramic binder after reaction (bottom).

## Coil technology

Coils used in accelerator magnets have relatively small bending radii and thus favor the Wind&React method. The superconducting Nb$_3$Sn phase in this case is formed after coil winding during high-temperature heat treatment. This technique requires that coil components (wedges, pole blocks, end parts, etc.) be capable of withstanding high-temperature heat treatment under compression. An optimization method for metallic end part design was developed and used at Fermilab [11]. Implementing the rapid prototyping technique enhanced the quality and reduced the time and the cost of end part development.

A critical innovation implemented at Fermilab to the coil fabrication process was using a liquid ceramic binder [12]. The ceramic binder improves the mechanical strength of cable insulation during coil winding and glues all the coil components after coil curing, thus simplifying coil handling, forming and measuring its cross-section before reaction. During the final coil heat treatment, the binder turns into small ceramic particles. These hard dielectric particles are excellent filler during coil impregnation with epoxy increasing the coil turn-to-turn electrical strength and its mechanical properties. These improvements simplified the coil fabrication process, increased its robustness, and reduced coil fabrication cost and time. Pictures of coils impregnated with ceramic binder after winding and curing and after reaction are shown in Fig. 5. The details of the baseline dipole and quadrupole coil technology are reported in [3], [5].

All the dipole and quadrupole coils were impregnated with CTD 101K epoxy to improve their mechanical and electrical properties. The radiation strength of the regular epoxy resin is quite low and that limits the lifetime of accelerator magnets operating in hard radiation environments. Fermilab is investigating some commercially available polyimide solutions [13] and new epoxy compounds to replace traditional epoxy as an impregnation material for Nb$_3$Sn coils.

## Mechanical structure and coil pre-load

Two quite different mechanical structures, one based on a thick stainless steel shell and the other one based on a stainless steel collar supported by stainless steel skin, were used in dipole and quadrupole models.

In the dipole structure the initial coil pre-stress of ~20 MPa and the magnet geometry control at room temperature is provided by two Al clamps. The final coil prestress of ~100-120 MPa at operating temperature, applied to reduce the radial and azimuthal turn motion under Lorentz forces, is created by the iron yoke, two clamps and a stainless steel skin.

The quadrupole mechanics involves coil initial pre-stress to ~30-50 MPa during collaring and then the final coil pre-stress to ~110-150 MPa by the stainless steel skin during assembly and cooling down to operating temperature. Control spacers prevent coil over-compression during yoking and skinning and increase the radial rigidity of the structure.

Axial coil pre-load and support in both dipole and quadrupole models is provided by thick end plates connected to the skin.

Nb$_3$Sn accelerator magnets with collar-based mechanical structures need a reliable collaring procedure for brittle Nb$_3$Sn coils [14]. The quadrupole coils collared with traditional quadrupole-style collars are usually compressed incrementally in the longitudinal direction. In order to limit the azimuthal stress gradient between adjacent sections, several passes are required to achieve the target coil prestress. The duration of the collaring procedure for this approach is proportional to the coil length and typically takes about one week per meter of coil. The maximum magnet length is also limited by the vertical space in a magnet assembly facility.

An alternative collaring method is based on a dipole-style collar. With this collar, the coils (dipole or quadrupole) are compressed simultaneously along their entire length, eliminating local stress gradients. This method lowers the risk of damage of brittle Nb$_3$Sn coils as well as significantly reduces the collaring time and makes it independent of coil length.

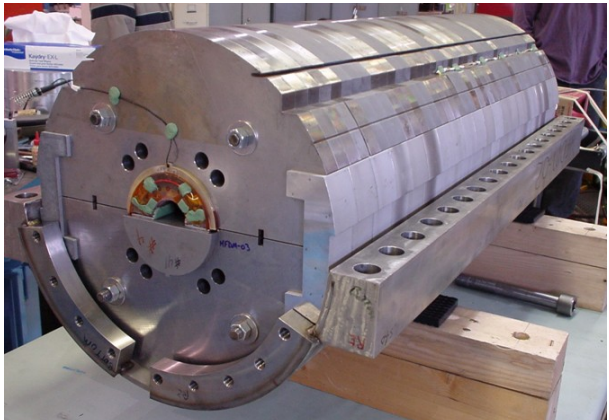

Figure 6: Dipole CTS (dipole mirror).

*Coil test structures*

Individual dipole and quadrupole coils were tested using special coil test structures (CTS) under operating conditions similar to those of real magnets, thus reducing the turnaround time of coil fabrication and evaluation, as well as material and labor costs. The dipole and quadrupole CTS use the same mechanical structures and assembly procedures as the corresponding complete magnets, and allow advanced instrumentation to be used.

The dipole CTS [15] is shown in Fig. 6. This structure is similar to the dipole structure of the HFDA series except that the iron yoke is split horizontally and one of the two coils is replaced with half-cylinder iron blocks. The coil inside the yoke is surrounded by bronze spacers. The transverse coil pre-stress and support is provided the same way as in the dipoles by a combination of the aluminum yoke clamps and the bolted stainless steel skin.

The quadrupole CTS [16] is shown in Fig. 7. It uses the iron yoke and skin of 90-mm quadrupoles of the TQC series. Three coils, collars and preload control spacers are replaced by iron blocks and spacers. This sub-assembly is installed in the standard TQC iron yoke and pre-compressed by a bolted stainless steel skin.

Axial coil pre-load and support in both dipole and quadrupole coil test structures is provided by two bolts in each thick end plate bolted to the skin.

## SHORT MODEL TEST RESULTS

Six short dipoles of the HFDA series and six dipole CTS of the HFDM series were built and tested during 2002-2006. This was the first series of nearly identical Nb$_3$Sn magnets which provided the first data on magnet quench performance and field quality and their reproducibility. In 2007-2010 seven quadrupole models of the TQC series and six quadrupole CTS of the TQM series were fabricated and tested, expanding and enriching the previous results and experience. In the course of the model magnet R&D phase the production time of short dipole and quadrupole models was reduced to 5-6 months per model, which is comparable with the production time of traditional NbTi dipole and quadrupole models.

The dipole models were tested in liquid helium normally at 4.5 K and some at lower temperatures. The quadrupole models were tested at 4.5 K, 1.9 K and intermediate temperatures.

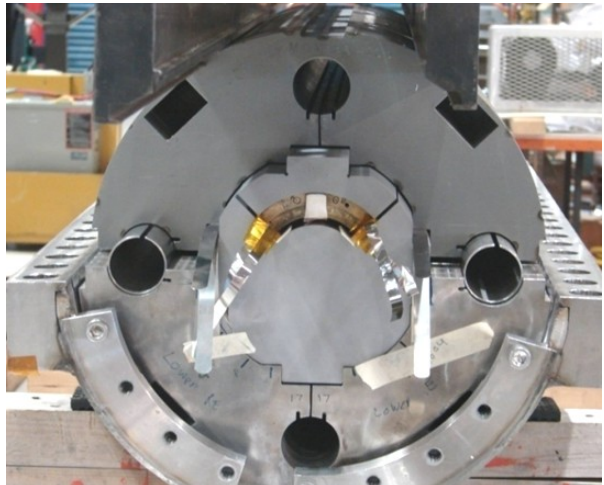

Figure 7: Quadrupole CTS (quadrupole mirror).

*Quench performance*

The first three dipole models, HFDA02-04, made of the MJR strand, were limited by flux jumps in the superconductor and reached 5-6 T or only 50-60% of their design field [17]. The last three dipole models HFDA05-07, made of the more stable 1-mm PIT-192 strand, reached $B_{max}$=9.4 T at 4.5 K and 10.2 T after cooling down to 2.2 K which corresponds to 100% of magnet short sample limit (SSL) at both temperatures. Fig. 8 shows the quench performance of the dipole models made of PIT strand. The maximum field reached by these models was ~10 T and was limited by the relatively low critical current density of the PIT strand. Nevertheless, these models clearly demonstrated that the developed Nb$_3$Sn coil technology and magnet mechanical structure are adequate for 10 T accelerator magnets.

A dipole coil made of high-$J_c$ RRP-108/127 strand and tested later in 2006 using the dipole test structure HFDM06 reached $B_{max}$= 11.4 T at 4.5 K (97% of SSL) confirming robustness of the developed dipole coil technology and mechanical structure (see next section).

The first quadrupole models TQC01a and TQC01b, made of the low-$J_c$ MJR strand, reached the nominal design field gradient of 200 T/m at 1.9 K [18].

Fig. 9 summarizes the quench performance of the quadrupole models TQC02Ea and TQC02Eb made of high-$J_c$ RRP-54/61 strand at 4.5 K. TQC02Ea was collared using traditional quadrupole collars and the multi-pass partial compression technique, whereas TQC02Eb was collared using the dipole-style collars. For comparison, magnet training data of TQS02a and TQS02c models utilizing the same set of coils and based on the alternative mechanical structure [19] are also presented. It can be seen that the quench performance of all the quadrupole models at 4.5 K was quite similar.

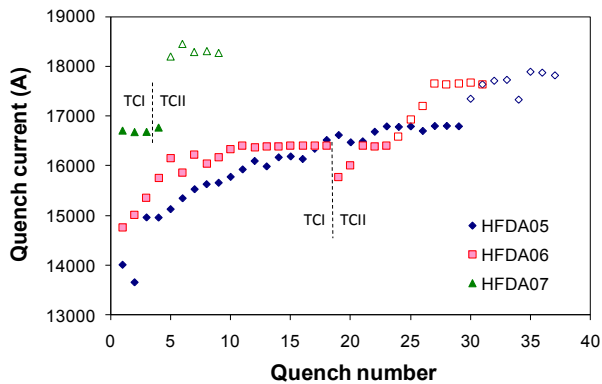

Figure 8: Dipole training quenches at 4.5 (solid markers) and 2.2 K (open markers) in thermal cycles TCI/TCII.

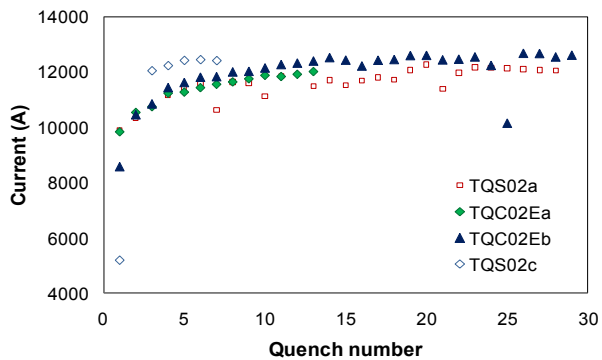

Figure 9: Quadrupole model training quenches at 4.5 K.

The maximum field gradient reached at 4.5 K in TQC models based on high-$J_c$ RRP-54/61 strand was 211 T/m or ~90% of magnet SSL. At 3 K it increased to 217 T/m and then at lower temperatures it reduced to ~ 200 T/m due to flux jumps in superconducting strands.

A TQC quadrupole model with coils made of RRP-108/127 strand is being assembled and will be tested with the goal of achieving the design field gradient of ~230 T/m at the nominal operation temperature of 1.9 K.

Both dipole and quadrupole short models demonstrated similar training performance including the relative level of the first quench, training duration and training memory after thermal cycling in spite of the significant difference in their structures and assembly techniques.

*Field quality*

The average values of geometrical harmonics in dipoles at 1.8 T and in quadrupoles at 45 T/m at the reference radius $R_{ref}$ corresponding to a half of the coil aperture are shown in Table 3. The values of the low order harmonics in both HFDA and TQC models are small, except for $b_3$ in HFDA and $a_4$ in TQC which are above one unit.

The standard deviations of normal and skew harmonics for HFDA dipole and TQC quadrupole models are shown in Fig. 10. The variation of skew harmonics in Nb$_3$Sn dipole and quadrupole models is quite close and still larger than in comparable dipole and quadrupole models based on traditional NbTi technology [20], [21]. The variation of normal harmonics is larger since it includes not only the coil component errors but also the adjustments of coil pre-stress shims. The reproducibility of both normal and skew harmonics in Nb$_3$Sn certainly can be improved by rising the tolerances of coil components, providing better coil alignment and reducing prestress variations.

Table 3: The average geometrical field harmonics for six dipole and five quadrupole models, $10^{-4}$.

| n | HFDA, $R_{ref}$=10 mm | | TQC, $R_{ref}$=22.5 mm | |
|---|---|---|---|---|
| | $a_n$ | $b_n$ | $a_n$ | $b_n$ |
| 2 | -0.37 | -0.15 | -0.25 | -0.09 |
| 3 | 0.55 | 2.06 | -0.45 | -0.97 |
| 4 | -0.73 | -0.06 | -1.46 | 0.28 |
| 5 | 0.17 | 0.60 | -0.25 | 0.97 |
| 6 | -0.04 | 0.00 | 0.06 | -0.02 |
| 7 | 0.01 | 0.20 | -0.08 | 0.10 |
| 9 | -0.02 | -0.05 | 0.04 | 0.04 |

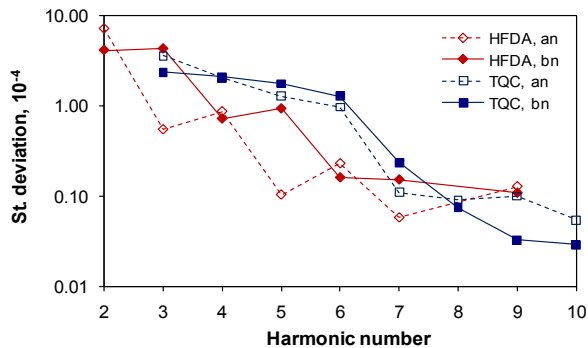

Figure 10: Normal $b_n$ and skew $a_n$ random field errors in Nb$_3$Sn dipole (HFDA) and quadrupole (TQC) models.

The coil magnetization, related to persistent currents in superconducting filaments and eddy currents in strands and cables, reduces the main field component (B in dipole and G in quadrupole) and affects the first allowed field harmonics - $b_3$ in dipoles and $b_6$ in quadrupoles.

The persistent current component is most important in the case of magnet operation with low ramp rates. It was large but reproducible in dipole and quadrupole models made of the same strand type [22], [23]. The higher strand $J_c$ or larger $d_{eff}$ proportionally changed the persistent current component of the magnet main field and the first allowed harmonics. In some dipole models with large flux jump activities in the coil, substantial erratic variations of sextupole field component at low fields were observed [24]. The superconductor magnetization theory and magnet experimental data suggest that the large persistent current effect and its variations observed in present Nb$_3$Sn accelerator magnets can be reduced by using strands with smaller sub-element size. A substantial fraction of the persistent current component can also be compensated using a passive correction based on thin iron strips [25].

The eddy current components depend on the current ramp rate, strand and cable twist pitches, transverse resistivity of the strand matrix, and interstrand resistance in the cable. The first three dipole models demonstrated a very small and reproducible eddy current effect due to large crossover resistances in the cable with the stainless steel core and the high resistivity (low RRR) of the strand matrix. The last three dipole models and all quadrupole models, all without stainless steel core in the cable and low matrix resistivity (high RRR), had large and non-reproducible eddy current components. This behavior was caused by the eddy currents in the cable due to large uncontrollable variations of cable interstrand resistance in coils. The above results suggest that the eddy current magnetization effect can be suppressed and well controlled by using cored cables and well-twisted strands.

Surprisingly, the decay and snapback effect, typical and quite strong in NbTi accelerator magnets, was not observed in either $Nb_3Sn$ dipole [26] or quadrupole [23] magnets. Studies of this effect will continue.

Table 4. Coil design features.

| CTS | Coil type | Strand type | $J_c$(12T, 4.2K), A/mm² | Filament $d_{eff}$, μm | Cable core | Cable insulation | Pole material |
|---|---|---|---|---|---|---|---|
| HFDM01 | DA05 | MJR-54/61 | 2200 | 100 | w/o core | Ceramic tape | Bronze |
| HFDM03 | DA12 | PIT-192 | 1600 | 50 | -"- | -"- | -"- |
| HFDM06 | DA19 | RRP-108/127 | 2100 | 70 | -"- | -"- | -"- |
| TQM01 | TQ19 | RRP-54/61 | 2800 | 70 | w/o core | S2-glass sleeve | Bronze |
| TQM02 | TQ17 | -"- | -"- | -"- | -"- | -"- | -"- |
| TQM03 | TQ34 | RRP-108/127 | 2500 | 50 | -"- | E-glass tape | Titanium |
| TQM04 | TQ35 | -"- | 2300 | -"- | 25 μm tape | S2-glass sleeve | -"- |

## $NB_3SN$ COIL STUDIES

Several issues were identified during the model magnet R&D which required experimental studies including the effect of conductor stability, cable core and insulation, coil pole materials, coil pre-stress. These and some other questions were studied and addressed by fabricating and testing series of dipole and quadrupole coils. The details of these studies are reported in [15]-[17], [27]. Coil design and fabrication features are summarized in Table 4.

Quench performance data of the dipole and quadrupole coils tested using the corresponding Coil Test Structures are plotted in Figs. 11 and 12.

The dipole coils made of different types of strand showed quite different training behavior. The coil made of the MJR strand with the largest value of $J_c \cdot d_{eff}$ and a relatively low RRR demonstrated erratic quench performance and large degradation of magnet quench current at 4.5 K. The PIT coil demonstrated stable training performance and reached its SSL at 4.5 K. At lower temperatures, it demonstrated the expected increase of the quench current.

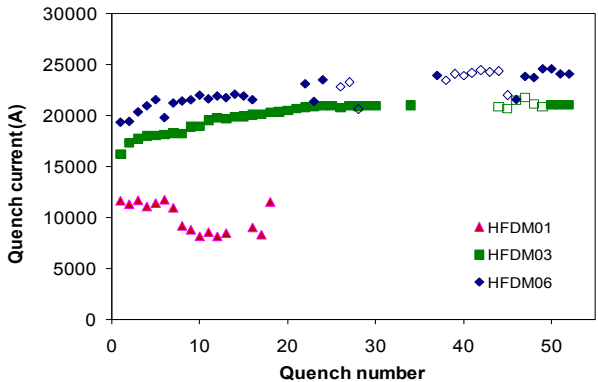

Figure 11: Dipole coil training quenches at 4.5 K (solid markers) and 2.2 K (open markers).

Table 5. Maximum pre-stress in the inner-layer pole turns.

| CTS | Coil pre-stress, MPa | |
|---|---|---|
| | 300 K | 4.5 K |
| TQM03a | 95 | 80 |
| TQM03b | 105 | 130 |
| TQM03c | 135 | 185 |

The RRP coil with reduced sub-element size reached the highest quench current, ~97% of its SSL limit, at 4.5 K. Noticeable variations of quench current on the current plateau pointed to mechanical or magnetic instabilities in the coil at high currents.

The quadrupole coils showed standard training behavior at 4.5 K with some variations of the first quench current, the number of training quenches and the maximum quench current. The training and ramp rate behaviors indicated that coils reached their SSL at 4.5 K. At 1.9 K the TQ coils (TQ17 and TQ19) made of RRP-54/61 strand with $d_{eff}$~70 μm showed some reduction of quench current and an erratic quench behavior which was observed also in the ramp rate measurements at the low current ramp rates. Meanwhile, coils TQ34 and TQ35, made of RRP-108/127 strand with $d_{eff}$~50 μm, showed the expected increase of quench current and regular ramp rate dependence at 1.9 K. After a few training quenches, these coils reached their SSL at 1.9 K.

To study the effect of pre-stress on the coil quench performance, coil TQ34 was assembled with three different warm and cold pre-stress values and tested three times using quadrupole CTS TQM03a/b/c. The values of maximum pre-stress in the inner-layer pole turns at room temperature and after cooling down are reported in Table 5. The TQM03a training data are shown in Fig. 12.

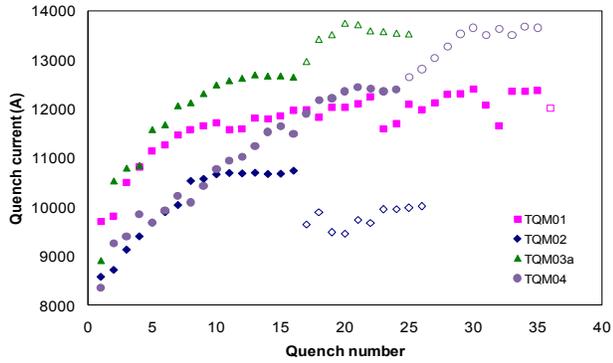

Figure 12: Quadrupole coil training quenches at 4.5 K (solid markers) and 1.9 K (open markers).

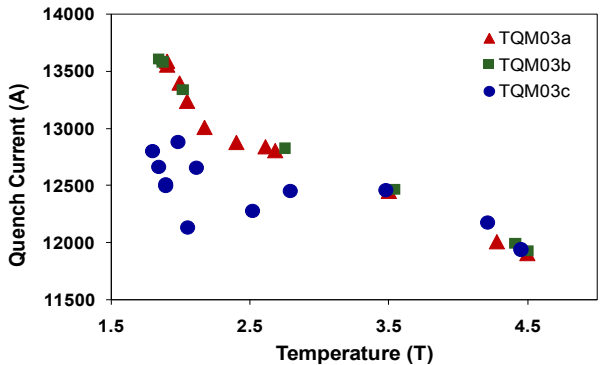

Figure 13: Coil TQ34 temperature dependence.

After re-assembly with higher pre-stress, TQM03b did not show any training and reached the same maximum quench currents both at 4.5 K and 1.9 K. TQM03c with the highest coil pre-stress also demonstrated good training memory at 4.5 K but unexpectedly low quench current increase and erratic quench behavior at 1.9 K.

The dependence of coil quench current on temperature for TQM03a/b/c is presented in Fig. 13. TQM03a and TQM03b showed stable and reproducible quenches over the entire temperature range from 1.9 to 4.5 K whereas TQM03c showed the same performance only above 3.5 K with most quenches below 3.5 K in the outer-layer blocks.

Analysis of the quench performance of the $Nb_3Sn$ dipole and quadrupole coils as well as the dipole and quadrupole models leads to the following practical conclusions:

a) The thin stainless steel core inside the cable does not degrade the coil training and maximum quench current but significantly reduces the sensitivity of the magnet quench current to the current ramp rate. It makes this approach an efficient means of suppressing eddy currents in the cable, which cause deterioration of field quality in $Nb_3Sn$ accelerator magnets during magnet ramping and unexpected magnet quenching during energy extraction.

b) The dipole and quadrupole coils with bronze and titanium pole parts and different cable insulation systems demonstrated similar quench performance. It confirms their compatibility with $Nb_3Sn$ coil technology for accelerator magnets.

c) The warm coil pre-stress up to 150 MPa and cold pre-stress up to 190 MPa do not cause any degradation of the coil critical current at 4.5 K. However, substantial flux jump instabilities at temperatures below 3 K were observed due to the possible local strand damage during coil fabrication and assembly, which led to a non-uniform transport current redistribution in strand cross-sections.

d) Flux jump instabilities in high-$J_c$ $Nb_3Sn$ strands with large $d_{eff}$ cause significant degradations of magnet quench performance. To suppress this effect in $Nb_3Sn$ accelerator magnets based on high-$J_c$ strand with $B_{nom}$ above 10 T, the value of $d_{eff}$ has to be less than 50 µm. To meet more strict field quality requirements at injection and provide conductor stability margin in the case of high coil pre-stress, the $d_{eff}$ should be even smaller, less than 20 µm.

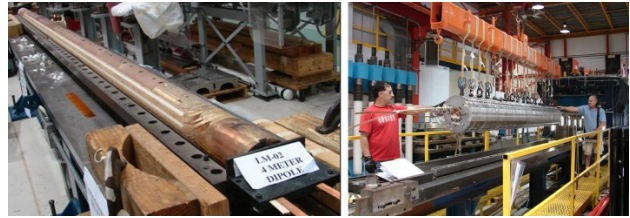

Figure 14: 4-m long $Nb_3Sn$ dipole coil (left) and LM02 cold mass (right).

## TECHNOLOGY SCALE UP

The technology scale up phase addresses the issues related to winding, curing, reaction, impregnation, and handling of long $Nb_3Sn$ coils, and long magnet assembly and performance due to the brittle nature of $Nb_3Sn$ superconductor. The scale-up was performed in several steps starting in 2007 with fabricating and testing a 2-m long dipole coil made of PIT $Nb_3Sn$ strand, which demonstrated stable and reproducible quench performance [22]. In 2008, the first 4-m long cos-theta dipole coil made of RRP-108/127 $Nb_3Sn$ strand was fabricated and tested [29]. The 4-m long $Nb_3Sn$ dipole coil and the 4-m long dipole CTS LM02 are shown in Fig. 14.

Training quenches of the 2-m long PIT coil (LM01) and the 4-m long RRP coil (LM02) at 4.5 K are shown in Fig. 15. The 2-m PIT coil after short training at 4.5 K reached its short sample limit and a field level of 10 T similar to the corresponding 1-m long PIT coil tested in dipole CTS HFDM03. The 4-m long dipole coil made of the high-$J_c$ RRP-108/127 strand, unlike its short version, was limited at 4.5 K by strong flux jump instabilities in the coil outer layer (perhaps caused by conductor damage during coil fabrication or CTS assembly). However, after suppressing them by heating the coil outer layer using quench heaters, it reached ~90% of its short sample limit at 4.5 K. The coil maximum quench current was limited by quenches in the inner-layer mid-plane turns caused by heaters. The described results are complemented by the results of $Nb_3Sn$ technology scale up performed by US-LARP by testing 4-m long racetrack coils [30] and recently the first 3.6-m long 90-mm quadrupole LQS01

[31]. The positive results of the Nb$_3$Sn technology scale up phase strengthen the high expectations for practical use of this technology in particle accelerators.

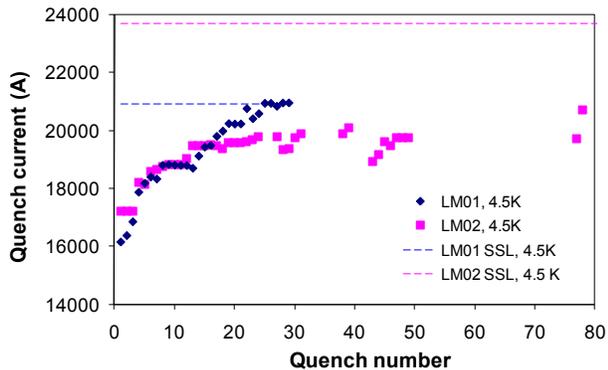

Figure 15: LM01 and LM02 training quenches at 4.5 K (markers) and short sample limits (dotted lines).

## CONCLUSIONS

Fermilab has been developing Nb$_3$Sn accelerator magnets over the past decade. The Nb$_3$Sn coil design and production experience includes ~20 dipole and ~35 quadrupole 1-m long coils as well as 2-m and 4-m long dipole coils, and 14 4-m long quadrupole coils fabricated completely at Fermilab or in collaboration with BNL and LBNL. The coil technology developed at Fermilab allowed reaching good reproducibility of the major coil parameters and short fabrication time. Two mechanical structures, one based on a thick stainless steel shell and the other based on a stainless steel collar supported by stainless steel skin, were developed and successfully tested. Two collaring techniques for brittle Nb$_3$Sn coils were also developed and experimentally demonstrated.

The robustness of the developed technologies was confirmed by handling and transportation of the short and long Nb$_3$Sn coils across the country, multiple coil re-assemblies in different mechanical structures and magnet tests without performance degradation.

The accelerator quality performance, including quench behavior and field quality, was reached in series of dipole and quadrupole models. The obtained results are not final, and there is room for their further improvement.

The advances in Nb$_3$Sn accelerator magnet technology during the past decade make it possible for the first time to consider Nb$_3$Sn magnets with nominal fields up to 12 T ($B_{max}$ up to 15 T) in present and future machines. To expand magnet operating fields up to 15 T, additional R&D effort will be required.

All the available experimental data show that superconductor properties are critical for magnet quench performance, field quality, protection, etc. Collaboration with materials groups in universities and industry on Nb$_3$Sn strand optimization is critical for the practical implementation of Nb$_3$Sn magnets in accelerators. The work on Nb$_3$Sn strand improvement with the goal of developing Nb$_3$Sn strands, which meet accelerator magnet specifications, has to be continued.

## ACKNOWLEDGMENTS
The author thanks the staff of Fermilab's Technical Division and colleagues at BNL, LBNL and CERN for their contributions to this effort.